\pdfoutput=1
\documentclass[a4paper,10pt]{article}
\usepackage{jheppub,amsmath,amsfonts,amssymb,xcolor,graphicx,hyperref}
\usepackage[utf8x]{inputenc}
\usepackage[section]{placeins}
\usepackage{cancel}
\usepackage{subfigure,color,dsfont,slashed,multirow}
\usepackage{fontenc}
\usepackage{pbox}
\usepackage{cancel}
\usepackage{setspace}
\usepackage{float}
\usepackage{mathtools,nccmath}%
\usepackage{etoolbox, xparse} 
\usepackage{array}
\usepackage[normalem]{ulem}
\usepackage{amstext}
\usepackage{soul}
\usepackage{amsmath,amssymb}
\usepackage{widetext}
\usepackage{cleveref}

\DeclarePairedDelimiterX{\abs}[1]\lvert\rvert{\ifblank{#1}{\,\cdot\,}{#1}}

\let\oldabs\abs
\def\abs{\futurelet\testchar\MaybeOptArgAbs}
\def\MaybeOptArgAbs{\ifx[\testchar\let\next\OptArgAbs
	\else \let\next\NoOptArgAbs\fi \next}
\def\OptArgAbs[#1]#2{\oldabs[#1]{#2}}
\def\NoOptArgAbs#1{\ifblank{#1}{\oldabs{}}{\oldabs[\big]{#1}}}

\DeclarePairedDelimiterX{\set}[1]\{\}{\setargs{#1}}
\NewDocumentCommand{\setargs}{>{\SplitArgument{1}{;}}m}
{\setargsaux#1}
\NewDocumentCommand{\setargsaux}{mm}
{\IfNoValueTF{#2}{#1}{\nonscript\,#1\nonscript\;\delimsize\vert\nonscript\:\allowbreak #2\nonscript\,}}
\let\oldset\set
\def\set{\futurelet\testchar\MaybeOptArgSet}
\def\MaybeOptArgSet{\ifx[\testchar \let\next\OptArgSet
	\else \let\next\NoOptArgSet \fi \next}
\def\OptArgSet[#1]#2{\oldset[#1]{#2}}
\def\NoOptArgSet#1{\OptArgSet[\big]{#1}}
\def\lsim{\raise0.3ex\hbox{$\;<$\kern-0.75em\raise-1.1ex\hbox{$\sim\;$}}}
\def\gsim{\raise0.3ex\hbox{$\;>$\kern-0.75em\raise-1.1ex\hbox{$\sim\;$}}}

\hypersetup{colorlinks=true,linkcolor=blue,citecolor=magenta,filecolor=magenta,
	urlcolor=cyan}
\newcolumntype{P}[1]{>{\centering\arraybackslash}p{#1}}

\hypersetup{bookmarks=true,unicode=true,pdftoolbar=true,pdfmenubar=true,
	pdffitwindow=false,pdfstartview={FitH},pdftitle={SecludedUMSSM},
	pdfauthor={~M.~Frank,Ya\c{s}ar Hi\c{c}y\i lmaz,Stefano Moretti,\"{O}.~\"{O}zdal}, pdfsubject={Subject},
	pdfcreator={Creator},pdfproducer={Producer}, pdfkeywords={Non-minimal 
		supersymmetry}{Beyond the Standard Model Physics}{SUSY},
	pdfnewwindow=true,colorlinks=true,
	linkcolor=blue,citecolor=magenta,filecolor=magenta,urlcolor=cyan}
\newcommand{\be}{\begin{equation}}
\newcommand{\ee}{\end{equation}}
\def\bsp#1\esp{\begin{split}#1\end{split}}
\renewcommand{\figureautorefname}{Fig.}

\def\sectionautorefname~#1\null{Sec.~(#1)\null}
\def\subsectionautorefname~#1\null{sub--Sec.~(#1)\null}
\def\figureautorefname~#1\null{Fig.~#1\null}
\def\tableautorefname~#1\null{Table~#1\null}
\def\equationautorefname~#1\null{Eq.~#1\null}

\newcommand{\mo}{\textsc{MicrOMEGAs}}

\newcommand{\spheno}{\textsc{SPheno}}

\mathchardef\mhyphen="2D

\def\mstone{m_{\tilde{t}_1}}
\def\msttwo{m_{\tilde{t}_2}}
\def\msbone{m_{\tilde{b}_1}}
\def\msbtwo{m_{\tilde{b}_2}}

\newcommand{\beq}{\begin{equation}}
\newcommand{\eeq}{\end{equation}}
\newcommand{\bea}{\begin{eqnarray}}
\newcommand{\eea}{\end{eqnarray}}

\newcommand{\mgut}{M_{{\rm GUT}}}

\newcommand{\WMSSM}{W_{{\rm MSSM}}}

\usepackage{jheppub} 

\usepackage[T1]{fontenc} 

\title{\boldmath $ t-b-\tau $ Yukawa Unification in Non-Holomorphic MSSM}

\author[a,b,1]{Ya\c{s}ar Hi\c{c}y\i lmaz,\note{Corresponding author.}}

\affiliation[a]{Department of Physics, Balikesir University, Balikesir 10145, Turkey}
\affiliation[b]{School of Physics and Astronomy, University of Southampton, Highfield, Southampton SO17 1BJ, United Kingdom}

\emailAdd{yasarhicyilmaz@balikesir.edu.tr}

\abstract{We show that in the CMSSM with the non-holomorphic soft SUSY breaking terms, the Yukawa coupling unification of the third family fermions at the GUT scale, called $ t-b-\tau $ Yukawa unification (YU), is possible under the recent collider and Dark Matter results. The YU parameter can also be found  $ R_{tb\tau} \approx 1 $, called perfect unification.  We find that the squark masses exceed 3 TeV while the stau can be considerably lighter. In the case of YU, the $ \tan\beta $ is in the interval $ [46,55] $. We obtain  bino-like dark matter (DM) of mass in the range of $ 0.6$ TeV $ \lesssim m_{\chi_{1}^{0}} \lesssim 1.3 $ TeV where the recent Dark Matter direct detection limits are also satisfied. We also identify  A-resonance solutions which reduce the relic abundance of LSP neutralino down to the ranges compatible with the current Planck measurements.}

\begin{document} 
\maketitle
\flushbottom

\section{Introduction}
\label{sec:intro}

Supersymmetry (SUSY) is a strong Beyond the Standard Model (BSM) theory that has various motivations such as resolution of the gauge hierarchy problem \cite{Barbieri:1987fn}, unification of the
gauge couplings \cite{Georgi:1974sy}, radiative electroweak symmetry breaking (REWSB)\cite{Higgs:1964pj,Englert:1964et}, dark
matter candidate under R-parity conservation, etc. However, on the experimental side, there have been not any clues from the SUSY partners of the Standard Model(SM)  particles at the LHC. This situation leads to a huge pressure on the SUSY models. In the minimal
supersymmetric extension of the SM (MSSM), the discovered 125 GeV SM-like Higgs boson \cite{Aad:2012tfa,Chatrchyan:2013lba} gives rise to issues related the fine-tuning problem. Similarly, the LHCb results for
the rare decays of B meson \cite{Amhis:2012bh,Aaij:2012nna} and the recent Dark Matter (DM) results from the astrophysical experiments \citep{Aghanim:2018eyx} have a significant impact on the parameter space of the SUSY models such as constrained MSSM (CMSSM) and non-universal Higgs mass models (NUHM) \cite{Roszkowski:2014wqa}. Moreover, in the MSSM, the experimental results that show significant deviation from SM predictions for the anomalous magnetic moment of muon $ (g − 2)_\mu $ requires very light smuon $ \tilde{\mu} $, as well as light electroweakinos, to get sufficient contribution from the SUSY sector, but the parameter space can not satisfy the LHC data in this region.

Nevertheless, the absence of any findings for SUSY can be translated into the presence of the non-minimal SUSY models in which the signals of a SUSY particle may be more complex to observe. Hence, such particles may also escape usual SUSY searches. There are too many non-minimal SUSY models such as U(1) extended MSSM (UMSSM) and Next-to-MSSM (NMSSM), etc. which have been extensively investigated in the literature and it has been concluded that such extended models are able to provide the solutions at the low energy scale that are in much better fit to the experimental results. Apart from the extensions of the MSSM with new particles and symmetries, a much simpler way to extend the MSSM can be obtained by adding non-holomorphic (NH) terms to the soft SUSY breaking (SSB) sector of the theory, so-called Non-Holomorphic MSSM (NHSSM). While the standard structure of the MSSM includes only holomorphic trilinear soft SUSY breaking terms \cite{Lykken:1996xt,Martin:1997ns,Chung:2003fi}, in a more general framework, the non-holomorphic supersymmetry breaking terms may also qualify as soft terms in the absence of a gauge singlet field \cite{Jack:1999ud,Martin:1999hc,Jack:1999fa}. Phenomenologically, such a framework can act differently from the standard MSSM. Thanks to the new trilinear interactions in NHSSM, a SM like CP even Higgs boson with mass ∼125 GeV can be achieved with relatively lighter squarks \cite{Chattopadhyay:2016ivr}. The NH terms, which alter the squark and electroweakino mass sectors, may also help to satisfy the experimental constraints from the rare B-decays and DM searches in the scenarios like phenomenological MSSM (pMSSM) \cite{Chattopadhyay:2016ivr}, constrained MSSM (CMSSM) \cite{Un:2014afa, Ross:2016pml, Ross:2017kjc} and minimal Gauge Mediated Supersymmetry Breaking (mGMSB)\cite{Chattopadhyay:2017qvh}. Moreover, a large SUSY contribution to $ (g − 2)_\mu $ can be possible even for heavy smuon \cite{Chattopadhyay:2016ivr}.

One of the significant implications of the SUSY SO(10) theories, such as the MSSM, is capability of the  the third generation ($ t-b-\tau $) Yukawa unification (YU) as well as gauge coupling unification at the grand unification (GUT) scale \cite{Ananthanarayan:1991xp,Shafi:1991rs}. In deed, the hierarchy of the third generation fermions (top quark, bottom quark and tau lepton) masses can be understood by YU. However, the CMSSM (or mSUGRA) with YU has been ruled for two reasons. First of them is that the exact YU with $ \mu > 0 $ and universal gaugino masses does not allow the neutralino dark matter scenario consistent with dark matter constraints \cite{Baer:2008jn, Baer:2008yd, Gogoladze:2009ug}. Secondly, in order to ensure a correct radiative electroweak symmetry breaking (REWSB) with YU, the MSSM Higgs soft supersymmetry breaking (SSB) masses must be split in such way that $ m_{H_u}^2<m_{H_d}^2 $  at the GUT scale \cite{Olechowski:1994gm, Matalliotakis:1994ft, Murayama:1995fn}. 

Whereas, in a constrained  NHSSM, these inconsistencies can be altered with the help of the additional NH terms. Here, it is important to note that YU is sensitive to low scale threshold corrections as well as the value of $ tan\beta $ \cite{Gogoladze:2011aa, Kolda:1995iw, Gogoladze:2009bn}, additionally, the NH soft terms can modify the threshold corrections to the fermion masses/Yukawa couplings \cite{Chattopadhyay:2018tqv}. So, it can be found the parameter space consistent with recent DM results as well as $ t-b-\tau$ YU. Further, the effects of the NH terms in Renormalization Group Equations (RGE) of the Higgs soft SSB masses \cite{Jack:1999fa, Ross:2016pml} may help to satisfy the REWSB with YU. With these motivations,  we probe in this work the third family  ($ t-b-\tau$) YU in NHSSM framework with fully universal boundary conditions at the GUT scale. We also test the low energy implications against all currently available data from the colliders and DM experiments, wich also includes direct detection searches.

The outline of the paper is as follows. We will briefly introduce the NHSSM in Section \ref{sec:model}. After summarising our scanning procedure and enforcing experimental constraints in Section \ref{sec:scan}, we present our results over the surviving parameter space and discuss the corresponding particle mass spectrum in Section \ref{sec:result}, including discussing DM implications. Finally, we summarise and conclude in Section \ref{sec:conc}.

\section{Model}
\label{sec:model}

 The superpotential in MSSM is given as

\begin{equation}
\WMSSM=\mu \hat{H}_{u}\hat{H}_{d}+Y_{u} \hat{Q} \hat{H}_{u} \hat{U}+Y_{d}\hat{Q}\hat{H}_{d} \hat{D}+Y_{e}\hat{L}\hat{H}_{d}\hat{E}
\label{MSSMW}
\end{equation}
where $\mu$ is the bilinear mixing term for the MSSM Higgs doublets
$H_{u}$ and $H_{d}$; $Q$ and $L$ denote the left handed squark and
lepton dublets, while $U,D,E$ stand for the right-handed u-type
squarks, d-type squarks and sleptons respectively. $Y_{u,d,e}$  are
the Yukawa couplings between the Higgs fields and the matter fields
shown as subscripts. {Higgsino mass term} $\mu$ is included in the SUSY preserving Lagrangian in MSSM, and hence it is allowed to be at any scale from
the electroweak (EW) scale to $\mgut$.  In addition to $\WMSSM$, the
soft SUSY breaking (SSB) Lagrangian is given below
\begin{equation*}\hspace{-5.7cm}
-\mathcal{L}^{soft}_{{\rm MSSM}}=m^{2}_{H_{u}}|H_{u}|^{2}+m^{2}_{H_{d}}|H_{d}|^{2}+m^{2}_{\tilde{Q}}|\tilde{Q}|^{2}+m^{2}_{\tilde{L}}|\tilde{L}|^{2}
\end{equation*}
\begin{equation*}
+m^{2}_{\tilde{U}}|\tilde{U}|^{2}+m^{2}_{\tilde{D}}|\tilde{D}|^{2}+m^{2}_{\tilde{E}}|\tilde{E}|^{2}+\sum_{a}M_{a}\lambda_{a}\lambda_{a}+(B\mu H_{u}H_{d}+{\rm h.c.})
\end{equation*}
\begin{equation}\hspace{-3.7cm}
+A_{u}Y_{u}\tilde{Q}H_{u}\tilde{U^{c}}+A_{d}Y_{d}\tilde{Q}H_{d}\tilde{D^{c}}+A_{e}Y_{e}\tilde{L}H_{d}\tilde{E^{c}}
\label{MSSM_SSB}
\end{equation}
where the field notation is as given before. $A_{u,d,e}$ are the
SSB terms for the trilinear scalar interactions, while $B$ is the
SSB bilinear mixing term for the MSSM Higgs fields.

The framework in this work differs from the standard MSSM only by introducing the contributions from additional soft supersymmetry breaking interactions, so called Non-Holomorphic (NH) terms as given below \cite{Jack:1999ud,Martin:1999hc,Jack:1999fa}

\begin{equation}
-\mathcal{L}^{NH}_{{\rm MSSM}}=\mu^\prime \tilde{H}_{u}\tilde{H}_{d}+A_{u}^\prime Y_{u}\tilde{Q}H_{d}\tilde{U^{c}}+A_{d}^\prime Y_{d}\tilde{Q}H_{u}\tilde{D^{c}}+A_{e}^\prime Y_{e}\tilde{L}H_{u}\tilde{E^{c}}+h.c.
\label{MSSM_NH}
\end{equation}
where $ A_{u,d,e}^{\prime} $ are the NH soft trilinear parameters corresponding to the up, down and lepton sectors, respectively and $ \mu^\prime $ is the NH soft Higgsino mass parameter. It can be noted that the new additional soft breaking sector consist of bare Higgsino mass terms and the trilinear couplings with "wrong" Higgs. In fact, such terms can be dangerous in the theories with gauge singlet fields as they break the SUSY hardly by regenerating the quadratic divergences \cite{Girardello:1981wz}. However, the NH SUSY breaking terms may be soft in the models, like MSSM, without any singlets \cite{Haber:2007dj,Hetherington:2001bk}.

The non-standard terms in Eqn. \ref{MSSM_NH} bring quite different phenomenology to the main sectors of the model at the low scale. Let us start with impact of NH terms on the scalar sector. The NH trilinear interactions containing $ A_{u,d,e}^{\prime} $ cause the modifications in tree-level sfermion mass-squared matrices of MSSM as follows.

\begin{eqnarray}
\label{eqn:sfermion_mass}
M_{\tilde{f}}^2=&\left(\begin{matrix}
m_{\tilde{f_L}}^2+(\frac{1}{2}-\frac{2}{3}\sin^2\theta_{W})m_Z^2\cos2\beta +m_f^2&\hspace{2mm}
-m_f(A_f-(\mu+A_f') C(\beta)) \\
-m_f (A_f-(\mu+A_f') C(\beta))  & \hspace{-6mm}
m_{\tilde{f_R}}^2+\frac{2}{3}\sin^2\theta_{W} m_Z^2 \cos2\beta +m_f^2 
\end{matrix} \right) \, \hspace{6mm}
\end{eqnarray}
where $m_{\tilde{f}_{L,R}}$ are the soft SUSY breaking masses for the
left- and the right-chiral sfermions, $m_f$ is the mass of the corresponding
fermion where $f=u,d,e $ stands for up-type quarks, down- type quarks and leptons respectively. $ C(\beta) = \cot\beta (\tan\beta)$ for the up-type squark (down-type squark, slepton) mass-squared matrices. $ A_{u,d,e}$ and $ A_{u,d,e}^{\prime}$ are 3x3 matrices in flavour space. As seen from  Eqn. \ref{eqn:sfermion_mass}, the NH trilinear terms with $ A_{f}^{\prime}$ appear in the off-diagonal elements of the sfermion mass squared matrices where $ \mu $ is replace by $ \mu+A_{f}^{\prime} $ and hence they can considerably change the mass of the sfermions by contributing to the L-R mixing. 

Such NH contributions to stop mass sector mainly effect the Higgs mass radiative corrections. In standard MSSM framework, the discovered Higgs boson with a mass of $ 125$ GeV \cite{Aad:2015zhl,Aad:2012tfa,Chatrchyan:2012ufa} requires large radiative corrections to the mass of the lighter neutral CP-even Higgs boson. This means that heavy scalar top quarks and large value of $ |A_t| $ on which the L-R mixing depends are needed in order to get large loop corrections for Higgs mass. The one loop correction to the lightest CP-even Higgs boson mass from top sector can be written as follows \cite{Djouadi:2005gj}

\begin{equation}
\label{higgs_loop}
\delta m_{h}^2= \frac{3 g^2_2 {\bar m}_t^4}{8 \pi^2 M_W^2} 
\left[\log\left(\frac{M_{S}^2}{{\bar m}_t^2}\right) + \frac{X_t^{2}}{2 M_{S}^2} 
\left(1 - \frac{X_t^{2}}{6M_{S}^2} \right) \right]. 
\end{equation}
where $ X_t = A_t-(\mu+A_t')\cot(\beta)$ is the left–right mixing term in NHSSM and $\frac{1}{2}(m_{\tilde{t}_1} + m_{\tilde{t}_2})$. $ {\bar m}_t $ is the running $\bar{MS} $ top quark mass to account
for the leading two–loop QCD and electroweak corrections in a RG improvement. It can be noted that this term corresponds the case of MSSM for $A_t'= 0 $. Unlike the standard MSSM framework, it is possible to satisfy the 125 GeV Higgs boson mass without having heavy stops or large $ |A_t| $ in NHSSM. As the YU can realize at the large $ \tan\beta $, the sbottom contributions will be significant in this work. It is important to emphasis that we take the loop contributions from all sectors into account. Moreover, even if the $ \mu^\prime $-term does not appear in the sfermion mass-squared matrices shown in Eqn. \ref{eqn:sfermion_mass}, it can contribute the Higgs masses through Higgsino loops at the higher order diagrams shown in Ref. \cite{Hollik:2014wea}. 

The another impact of the non-holomorphic terms is on the masses of the electroweakinos. In fact, only $ \mu^\prime $-term changes the neutralino and chargino mass matrices as shown below.

\begin{eqnarray}
\label{neutralino_mass}
 M_{\widetilde{\chi^0}}=&\left(\begin{matrix}
                        M_1 & 0 & -M_Z\cos\beta \sin\theta_W & M_Z\sin\beta \sin\theta_W \\
                        0 & M_2 & M_Z\cos\beta \cos\theta_W & -M_Z\sin\beta \cos\theta_W \\
                       -M_Z\cos\beta \sin\theta_W & M_Z\cos\beta \cos\theta_W & 0 & -(\mu+\mu^\prime)\\
                        M_Z\sin\beta \sin\theta_W & -M_Z\sin\beta \cos\theta_W & -(\mu+\mu^\prime) & 0
                                           \end{matrix} \right).\hspace{6mm}
\end{eqnarray}
\begin{eqnarray}
\label{chargino_mass}
  M_{\widetilde{\chi^{\pm}}}=&\left(\begin{matrix}
                                 M_2 & \sqrt{2} M_W\sin\beta \\
                                 \sqrt{2} M_W\cos\beta & (\mu+\mu^\prime)\\
                                \end{matrix}\right).
\end{eqnarray}
where $M_{\tilde{\chi}^{0}}$ is mass matrix for the neutralinos in
the basis ($\tilde{B},\tilde{W}^{0},
\tilde{H}_{d}^{0},\tilde{H}_{u}^{0}$) and ($\tilde{B},\tilde{W}^{0},
\tilde{H}_{d}^{0},\tilde{H}_{u}^{0}$), while
$M_{\tilde{\chi}^{\pm}}$ is for the charginos in the basis
$(\tilde{W}^{-},\tilde{H}_{d}^{-})$ and
$(\tilde{W}^{+},\tilde{H}_{u}^{+})$.


\begin{figure}[t]
     \begin{center}
        {%
            \label{fig:loopdiagsA}
            \includegraphics[width=0.40\textwidth]{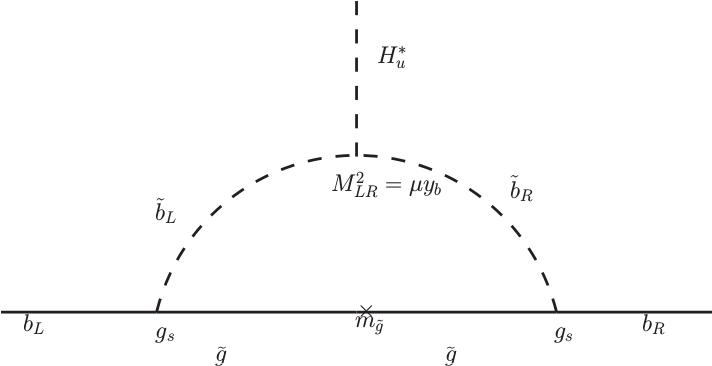}
        }%
\hskip 30pt
        {%
           \label{fig:loopdiagsB}
           \includegraphics[width=0.40\textwidth]{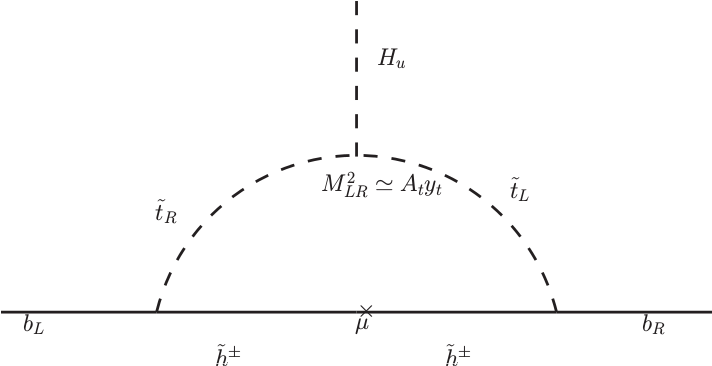}
        }%
        \\
\vspace{0.75cm}
        {%
           \label{fig:loopdiagsC}
           \includegraphics[width=0.40\textwidth]{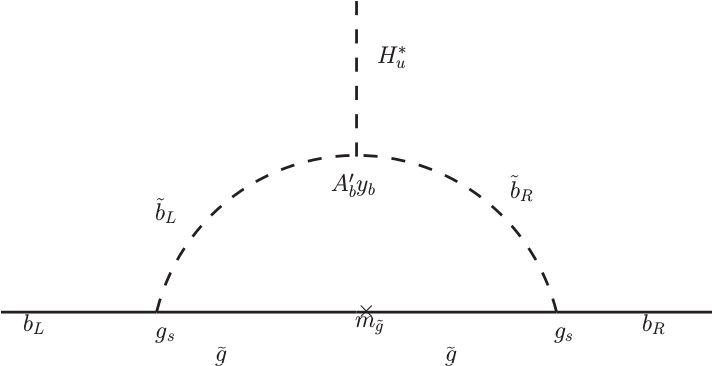}
        }%
\hskip 30pt
        {%
           \label{fig:loopdiagsD}
           \includegraphics[width=0.40\textwidth]{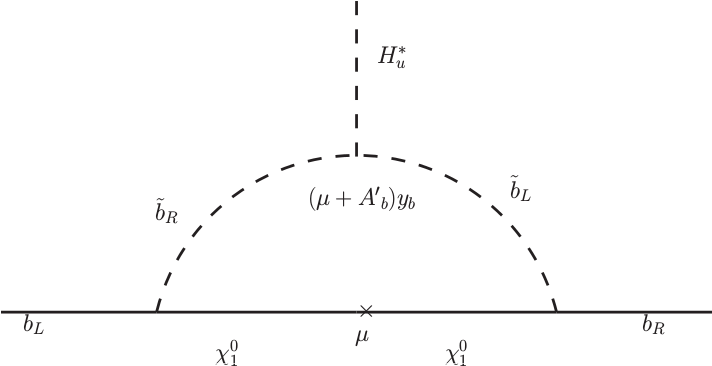}
        }%
        \caption{The main diagrams for the one loop contribution to the bottom quark mass in MSSM (top planes) and the NHSSM (bottom planes) \cite{Chattopadhyay:2018tqv}}
\label{fig:bb}
\end{center}
\end{figure}

Undoubtedly, the NH terms shown in Eqn. \ref{MSSM_NH} can also change the contribution of MSSM to masses of quarks and leptons at the loop level since they introduce new Yukawa interactions between the sfermion and sfermion through reverse Higgs in addition to standard Yukawa interactions shown in MSSM superpotential. Fig. \ref{fig:bb} shows the one loop diagrams for sbottom and stop contributions to the bottom quark mass involving change of gluino, charginos and neutralinos. In the MSSM, the interactions between sbottom (stop) and Higgs fields is $ \tilde{b}\tilde{b}H_d $ ($ \tilde{t}\tilde{t}H_u $) while there is an additional interaction as $ \tilde{b}\tilde{b}H_u $  ($\tilde{t}\tilde{t}H_d $) in NHSSM. The $ \mu^\prime  $ term also enters these diagrams through chargino/neutralino masses shown in Eqns. \ref{neutralino_mass} and \ref{chargino_mass}. In the case of the MSSM, the main contributions to bottom mass(or $ y_b $) arise from gluino-sbottom and chargino-stop loops whose diagrams are shown in the top planes of Fig. \ref{fig:bb} and are written by \cite{Hall:1993gn, Hempfling:1993kv, Pierce:1996zz}
\begin{eqnarray}
{\Delta m_b^{(\tilde g)}}_{\rm MSSM}&=&\frac{2\alpha_3}{3\pi}m_{\tilde g} \mu y_b \frac{v_u}{\sqrt 2} I( \msbone^2, \msbtwo^2, m_{\tilde g}^2), \nonumber \\
{\Delta m_b^{{{\tilde h}^+}}}_{\rm MSSM} &=& \frac{y_t y_b}{16 \pi^2} \mu A_t y_t \frac{v_u}{\sqrt 2} I( \mstone^2, \msttwo^2, \mu^2) \, , 
\label{MSSM_loop}
\end{eqnarray}
\noindent
where the loop integral $I(a,b,c)$ is given by
\begin{equation}
I(a,b,c)=-\frac{a b~\ln(a/b)+bc~\ln(b/c)+c a~\ln(c/a)}{(a-b)(b-c)(c-a)}.
\end{equation}  
\noindent
In the NHSSM, there are additional contributions from gluino-sbottom and neutralino-sbottom loops whose diagrams are shown in the bottom planes of Fig. \ref{fig:bb} and for $\mu^\prime = 0$ can be given by 
\begin{eqnarray}
  {\Delta m_b^{(\tilde g)}}_{\rm NHSSM}&=&\frac{2\alpha_3}{3\pi}m_{\tilde g}
  {A_b^\prime} y_b 
\frac{v_u}{\sqrt 2} I( \msbone^2, \msbtwo^2, m_{\tilde g}^2), \nonumber \\
     { \Delta m_b^{{{\tilde h}^0}}}_{\rm NHSSM} &=& \frac{y_b^2}{16 \pi^2} \mu
     {(\mu+{A_b^\prime}) y_b} \frac{v_u}
{\sqrt 2} I( \msbone^2, \msbtwo^2, \mu^2). 
\label{NHSSM_loop}
\end{eqnarray}
and for $\mu^\prime \neq 0$, all the loop contributions to bottom mass approximately can be written by \cite{Chattopadhyay:2018tqv}
\begin{eqnarray}
\Delta{m_b} & \approx & \frac{y_b v_d}{\sqrt 2}
\Big[\frac{y_t^2}{16 \pi^2}\mu A_t I( \mstone^2, \msttwo^2, (\mu^2+{\mu^{\prime}}^2)) \tan\beta +
   \frac{2\alpha_3}{3\pi}m_{\tilde g} (\mu + A_b^\prime)
  I( \msbone^2, \msbtwo^2, m_{\tilde g}^2)\tan\beta
\nonumber \\
 &+& \frac{y_b^2}{16 \pi^2}\mu (\mu+A_b^\prime) I( \msbone^2, \msbtwo^2, 
(\mu^2+{\mu^{\prime}}^2)) \tan\beta \Big].
\label{cont_b} 
\end{eqnarray}

Such additional contributions to quark and lepton masses in NHSSM are translated into the modifications in the threshold corrections to their Yukawa couplings, which are crucial in realizing Yukawa Unification consistent with the observed quark and lepton masses for the third family. The finite correction to Yukawa coupling of the b-quark is needed especially to be appropriately large and negative \cite{Gogoladze:2010fu}. With the help of the new contributions in Eqn. \ref{cont_b}, this condition, especially in the constrained version of NHSSM, may be satisfied with the parameter space which provides compatible solutions with the recent LHC and DM results.

Moreover, the NH terms $ \mu^\prime $, $ A_{u}^\prime $, $ A_{d}^\prime $ and $ A_{e}^\prime $ can enter the RGEs of the two Higgs field soft breaking terms, $ m_{H_u}^2 $ and $ m_{H_d}^2 $ whose one loop $ \beta $ functions are as shown below\cite{Ross:2016pml}.
\begin{align} 
\beta_{m_{H_d}^2}^{(1)} & =  
-\frac{6}{5} g_{1}^{2} {\mu'}^{2} -6 g_{2}^{2} {\mu'}^{2} -\frac{6}{5} g_{1}^{2} |M_1|^2 -6 g_{2}^{2} |M_2|^2 - g_1\sigma\nonumber \\ 
 &+6 \mbox{Tr}\Big({{A^{'}_u}  {A^{'}_{u}}^{\dagger}}\Big) +6 m_{H_d}^2 \mbox{Tr}\Big({Y_d  Y_{d}^{\dagger}}\Big) +2 m_{H_d}^2 \mbox{Tr}\Big({Y_e  Y_{e}^{\dagger}}\Big) \nonumber \\ 
 &+6 \mbox{Tr}\Big({A_d^*  A_{d}^{T}}\Big)  +2 \mbox{Tr}\Big({A_e^* A_{e}^{T}}\Big) +6 \mbox{Tr}\Big({m_d^2  Y_d  Y_{d}^{\dagger}}\Big) \nonumber \\ 
 &+2 \mbox{Tr}\Big({m_e^2  Y_e  Y_{e}^{\dagger}}\Big) +2 \mbox{Tr}\Big({m_l^2  Y_{e}^{\dagger}  Y_e}\Big) +6 \mbox{Tr}\Big({m_q^2  Y_{d}^{\dagger}  Y_d}\Big) \nonumber \\
\beta_{m_{H_u}^2}^{(1)} & =  
-\frac{6}{5} g_{1}^{2} {\mu'}^{2} -6 g_{2}^{2} {\mu'}^{2} -\frac{6}{5} g_{1}^{2} |M_1|^2 -6 g_{2}^{2} |M_2|^2 + g_1\sigma\nonumber \\ 
 &+6 \mbox{Tr}\Big({{A^{'}_d}  {A^{'}_{d}}^{\dagger}}\Big) +2 \mbox{Tr}\Big({{A^{'}_e}  {A^{'}_{e}}^{\dagger}}\Big) +6 m_{H_u}^2 \mbox{Tr}\Big({Y_u  Y_{u}^{\dagger}}\Big) \nonumber \\ 
 &+6 \mbox{Tr}\Big({A_u^*  A_{u}^{T}}\Big) +6 \mbox{Tr}\Big({m_q^2  Y_{u}^{\dagger}  Y_u}\Big) +6 \mbox{Tr}\Big({m_u^2  Y_u  Y_{u}^{\dagger}}\Big).
\label{beta_mHu}
\end{align} 
In the MSSM framework, it is not possible to achieve the $ t-b-\tau $ YU with universal scalar masses at the GUT scale, hence, the Higgs field soft breaking terms can not ensure the electroweak symmetry breaking condition, which is that $ m_{H_u}^2 $ must beless than $m_{H_d}^2 $ negative at the low scale \cite{Murayama:1995fn,Baer:2009ie}. Considering the different contributions of the NH soft trilinear parameters to the RGEs of the each Higgs field soft breaking terms in Eqn. \ref{beta_mHu}, unlike the MSSM, it can be possible that the Higgs field soft breaking terms which are degenerate at $ M_{GUT} $ split at the low scale to allow for an appropriate radiative breakdown of electroweak symmetry (REWSB).

\section{Scanning Procedure and Experimental Constraints}
\label{sec:scan}

In our work, we have employed the $\spheno$ (version 4.0.3) package \cite{Porod:2003um} obtained with {\tt SARAH} (version 4.14.3) \cite{Staub:2008uz}. In this code, all gauge and Yukawa couplings in the NHSSM are evolved from the EW scale to the GUT scale that is assigned by the condition of gauge coupling unification, described as $ g_{1}=g_{2}$ . However, $ g_{3} $ is allowed to have a small deviation from the unification condition, since it has the largest threshold corrections at the GUT scale \cite{Hisano:1992jj}. After that, the whole mass spectrum is calculated by evaluating all SSB parameters along with gauge and Yukawa couplings back to the EW  scale. These bottom-up and top-down processes are carried out by running the RGEs and in the latter, boundary conditions given at $ M_{\rm GUT} $ scale are used. We assume the universal boundary conditions at the GUT scale:
\begin{eqnarray}
 M_1 &=& M_2 = M_3 \equiv M_{1/2}\nonumber \\
 m_{H_d}^2 &=& m_{H_u}^2 \equiv m_0^2  \nonumber \\
 m_e^2 &=& m_d^2 = m_u^2 = m_l^2 = m_q^2 \equiv \textbf{1} m_0^2 \nonumber \\
 A_i &=& A_0 Y_i \nonumber \\
 A_i^\prime &=& A_0 Y_i^\prime \nonumber
\end{eqnarray}
where $ m_{0} $ is the universal SSB mass term for the matter scalars while  $ M_{1/2} $ are the universal SSB mass terms of the gauginos at the GUT scale. Besides, $ A_0 $ is the SSB trilinear coupling  and $ A_0^\prime $ is the NH SSB trilinear coupling. $ i = u, d, e $ stands for up-type quarks, down-type quarks and leptons respectively. In the numerical analysis of our work, we have performed random scans over the parameter space of the NHSSM shown in Table \ref{paramSP}, where $ \tan\beta $ is the ratio of the VEVs of the MSSM Higgs doublets and $ \mu^\prime $ is the NH soft Higgsino mass parameter.

\begin{table}[H]
	\centering
	\begin{tabular}{c|c||c|c}
		Parameter  & Scanned range & Parameter      & Scanned range \\
		\hline
		$m_0$ & $[0, 6]$ TeV     & sign  $\mu$    & $ + 1 $ \\
		$M_{1/2}$        & $[0, 6]$ TeV & $A_0^\prime$ & $[-15, 15]$ TeV\\
		$\tan\beta$ & $[1, 60]$   & $\mu^\prime$    & $[-5, 5]$ TeV\\
		$A_0$    & $[-5, 5]$ TeV  &  \\
	\end{tabular}
	\caption{Scanned parameter space.}
	\label{paramSP}
\end{table}
In order to scan the parameter space efficiently, we use the Metropolis-Hasting algorithm \cite{Belanger:2009ti}. Then, we implement Higgs boson and sparticle mass bounds \cite{Chatrchyan:2012xdj,Tanabashi:2018oca} and the constraints from rare $B$-meson decays such as $ {\rm BR}(B \rightarrow X_{s} \gamma) $ \cite{Amhis:2012bh}, $ {\rm BR}(B_s \rightarrow \mu^+ \mu^-) $ \cite{Aaij:2012nna} and $ {\rm BR}(B_u\rightarrow\tau \nu_{\tau}) $ \cite{Asner:2010qj}. We also require that the predicted relic density of the neutralino LSP agrees within 5$ \sigma $ with the  recent Planck  results \cite{Aghanim:2018eyx}. The relic density of the LSP and scattering cross sections for direct detection experiments are calculated with $\mo$ (version 5.0.9) \cite{Belanger:2018mqt}. The experimental constraints can be summarised as follows:
\begin{equation}
\setstretch{1.8}
\begin{array}{l}
m_h  = 122-128~{\rm GeV},
\\
m_{\tilde{g}} \geq 2.0~{\rm TeV},
\\
0.8\times 10^{-9} \leq{\rm BR}(B_s \rightarrow \mu^+ \mu^-)
\leq 6.2 \times10^{-9} \;(2\sigma~{\rm tolerance}),
\\
m_{\tilde{\chi}_{1}^{\pm}} \geq 103.5~{\rm GeV}, \\
m_{\tilde{\tau}} \geq 105~{\rm GeV}, \\
2.99 \times 10^{-4} \leq
{\rm BR}(B \rightarrow X_{s} \gamma)
\leq 3.87 \times 10^{-4} \; (2\sigma~{\rm tolerance}),
\\
0.15 \leq \dfrac{
	{\rm BR}(B_u\rightarrow\tau \nu_{\tau})_{\rm NHSSM}}
{{\rm BR}(B_u\rightarrow \tau \nu_{\tau})_{\rm SM}}
\leq 2.41 \; (3\sigma~{\rm tolerance}), \\
0.114 \leq \Omega_{{\rm CDM}}h^{2} \leq 0.126~(5\sigma~{\rm tolerance}).
\label{constraints}
\end{array}
\end{equation}
We have also applied the the colour and charge breaking (CCB) constraint in NHSSM shown in Ref. \cite{Beuria:2017gtf}. 

Additionally, we implement a constraint on the parameter that quantifies the $ t-b-\tau $ YU as shown below. 

\begin{equation}
R_{tb\tau} = \frac{Max(y_t,y_b,y_{\tau})}{Min(y_t,y_b,y_{\tau})},
\end{equation}
where $ R_{tb\tau} \leq 1.1 $ means that the solution are compatible with $ t-b-\tau $ YU \cite{Baer:2012cp}. $ R_{tb\tau} = 1 $ donates perfect $ t-b-\tau $ YU. As mentioned in Section \ref{sec:model}, this condition can not be satisfied in the standard CMSSM. The following  list  summarises the relation between colours and constraints imposed in our forthcoming plots.

\begin{itemize}
	\item Grey: Radiative EWSB (REWSB) and  neutralino LSP.
	\item Red: The subset of grey  plus  Higgs boson mass and coupling constraints, SUSY particle mass bounds and EWPT requirements.
	\item Green:  The subset of red  plus  $B$-physics constraints.
	\item Blue: The subset of green  plus Planck constraints on the relic abundance of the neutralino LSP (within $5\sigma$).
	\item Black: The subset of blue  plus $ t-b-\tau $ Yukawa Unification constraint, $ R_{tb\tau} \leq 1.1 $ 
\end{itemize}

\newpage
\section{Results}
\label{sec:result}

\begin{figure}[!t]
	\centering
	\includegraphics[width=0.48\linewidth]{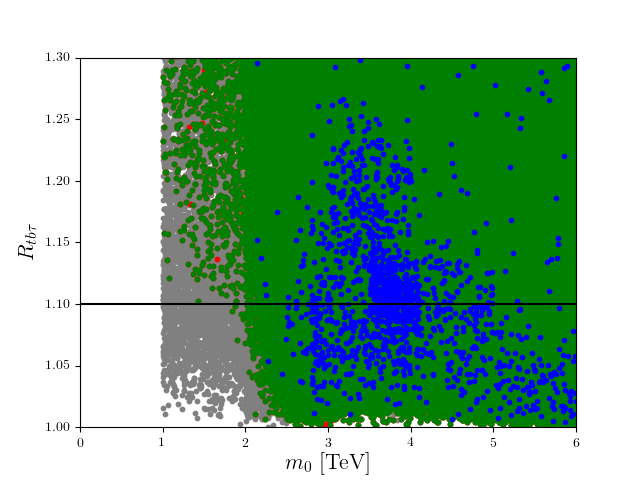}
	\includegraphics[width=0.48\linewidth]{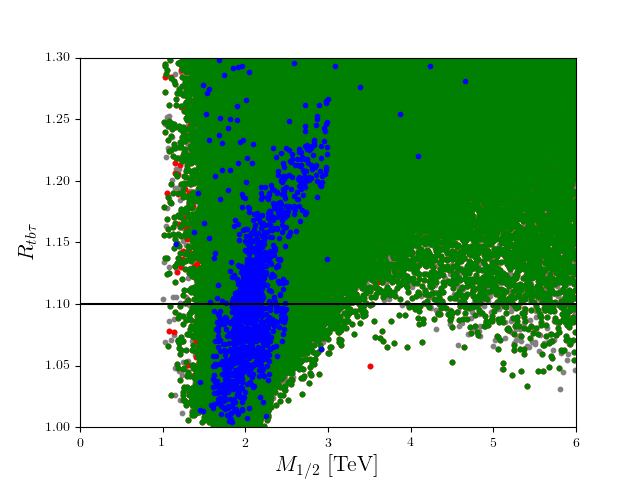}
	\includegraphics[width=0.48\linewidth]{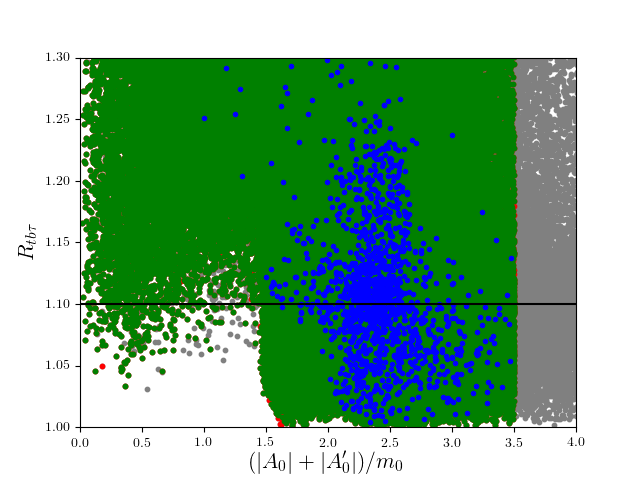}
	\includegraphics[width=0.48\linewidth]{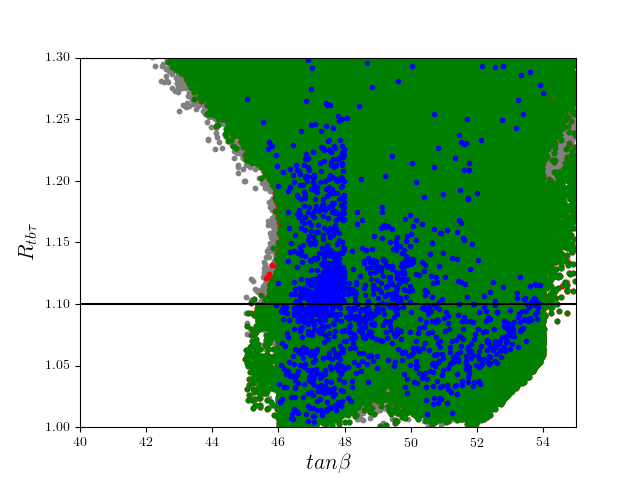}
	\includegraphics[width=0.48\linewidth]{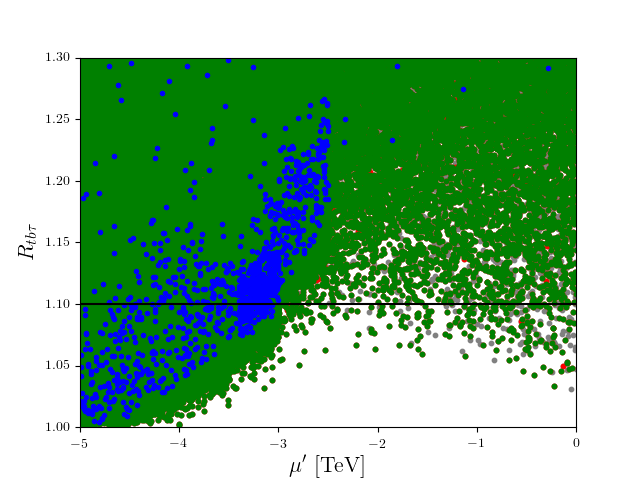} 
	\caption{The fundamental parameter space versus $ t-b-\tau $ YU ($ R_{tb\tau} $) over the following planes: $ m_0-R_{tb\tau} $ 
		(top left), $ M_{1/2}-R_{tb\tau} $ (top right), $ (|A_0|+|A_0^{\prime}|^2)/m_0-R_{tb\tau} $ 
		(middle left), $ tan\beta-R_{tb\tau} $ (middle right) and $ \mu^\prime-R_{tb\tau} $ (bottom). Our colour convention is as listed at the end of Section \ref{sec:scan}. The solid line indicates the region consistent with YU with $ R_{tb\tau} \leq 1.1$}
	\label{fig:param1}
\end{figure}
In this section, we will first present the regions in the parameter space that are compatible with $ t-b-\tau $ YU in Fig. \ref{fig:param1} with $ m_0-R_{tb\tau} $ (top left), $ M_{1/2}-R_{tb\tau} $ (top right), $(|A_0|+|A_0^{\prime}|^2)/m_0-R_{tb\tau} $  (middle left), $ tan\beta-R_{tb\tau} $ (middle right) and $ \mu^\prime-R_{tb\tau} $ (bottom) planes. The colour convention is as listed at the end of Section \ref{sec:scan} and the solid line indicates the region consistent with YU with $ R_{tb\tau} \leq 1.1$. According to top panels, the $ t-b-\tau $ YU allows a wide range for $ m_0 $, namely $ 2 \ {\rm TeV} \lesssim m_0 \lesssim 6 \ {\rm TeV}$ while $ M_{1/2} $ lies in the range of $ 1.5 \ {\rm TeV} \lesssim M_{1/2} \lesssim 3 \ {\rm TeV}$. Essentially, perfect YU is realized for $ m_0 > 4 \ {\rm TeV} $ and $ M_{1/2} \approx 2 \ {\rm TeV} $. In the middle left panel, we show the ratio $ (|A_0|+|A_0^{\prime}|^2)/m_0 $ which must be smaller than $ 3.5 $ to satisfy CCB condition in \cite{Beuria:2017gtf}. We observe that the main contribution to this ratio comes from the NH SSB trilinear coupling $ A_0^{\prime} $ in the range $ -15 \ {\rm TeV} \lesssim A_0^{\prime} \lesssim -8 \ {\rm TeV}$, since it is needed to be large and negative in order to enhance the effect of the last term in Eqn. \ref{cont_b}. As seen from the middle right panel, $ tan\beta $ should be in the interval $ [45,55] $. In the bottom panel, it can be seen that the NH soft Higgsino mass parameter $ \mu^\prime $ should be negative in the range of $ -5 \ {\rm TeV} \lesssim \mu^{\prime} \lesssim -3 \ {\rm TeV}$. Moreover, the perfect $ t-b-\tau $ YU can be actualized for $ \mu^\prime < -4.5 \ {\rm TeV} $.

\begin{figure}[!t]
	\centering
	\includegraphics[width=0.48\linewidth]{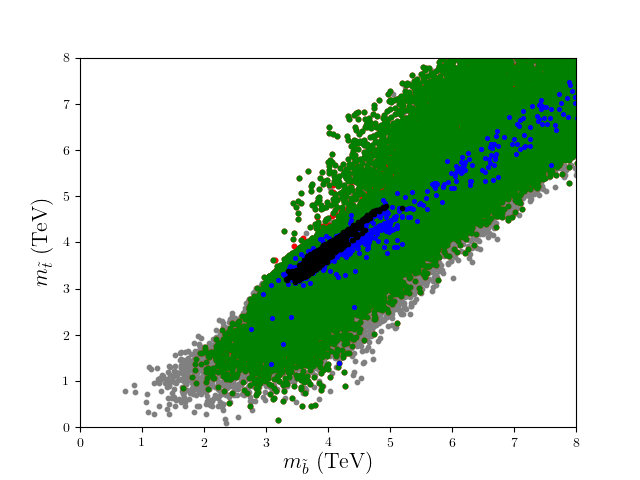}
	\includegraphics[width=0.48\linewidth]{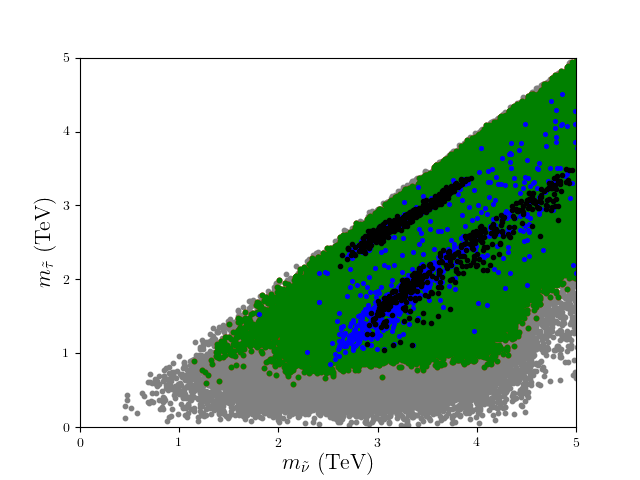}
	\caption{The mass spectrum of SUSY states over the following planes:  $(m_{\tilde{b}},m_{\tilde{t}})$ (left) and $(m_{\tilde{\nu}},m_{\tilde{\tau}})$ (right). Our colour convention is as listed at the end of Section \ref{sec:scan}.}
	\label{fig:spar}
\end{figure}

Fig. \ref{fig:spar} displays the the mass spectrum of SUSY particles in $(m_{\tilde{b}},m_{\tilde{t}})$ (left) and $(m_{\tilde{\nu}},m_{\tilde{\tau}})$ (right) planes.The colour convention is as listed at the end of Section \ref{sec:scan}. In addition to the colours in the Fig. \ref{fig:param1}, we have also added the black colour in which the points are compatible with $ t-b-\tau$ YU. The left panel shows that stop and sbottom masses should be $ 3 \ {\rm TeV} \lesssim m_{\tilde{b}},m_{\tilde{t}} \lesssim 5 \ {\rm TeV}$.  These mass ranges are not quite heavy and they can be probed in the future collider searches. As can be seen from the right panel in Fig. \ref{fig:spar}, stau $ \tilde{\tau}  $ can be as light as 1 TeV, compatible with $ t-b-\tau $ YU while sneutrino can be cover a heavier range, $ 2.5 \ {\rm TeV} \lesssim m_{\tilde{b}},m_{\tilde{t}} \lesssim 5 \ {\rm TeV}$.  According to our results, we also obtain the gluino mass with $ m_{\tilde{g}} \gtrsim 3 \ {\rm TeV} $.

\begin{figure}[!t]
 	\centering
 	\includegraphics[width=0.48\linewidth]{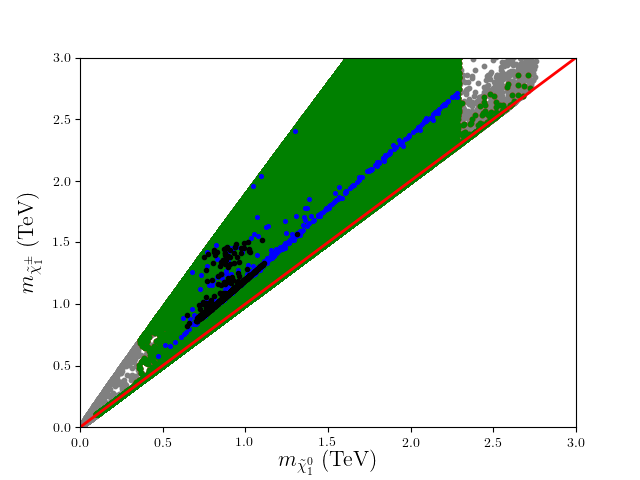}
 	\includegraphics[width=0.48\linewidth]{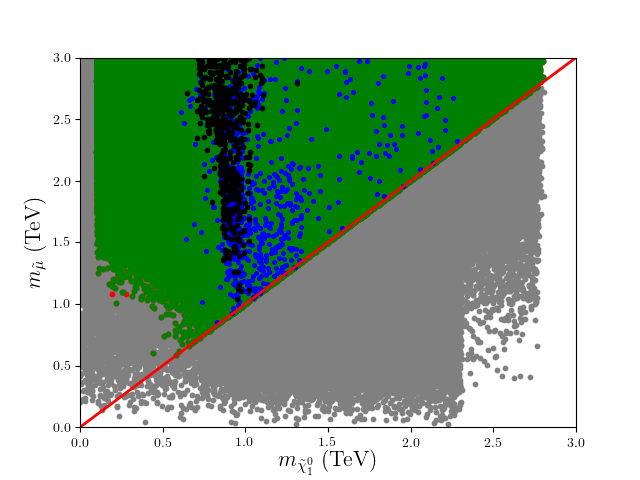}
 	\includegraphics[width=0.48\linewidth]{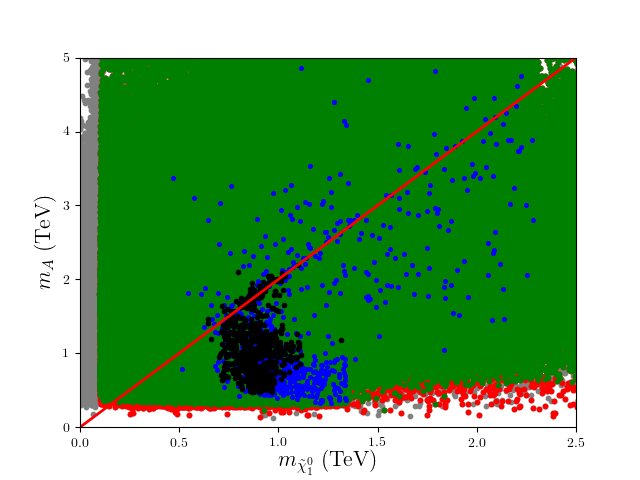}
 	\includegraphics[width=0.48\linewidth]{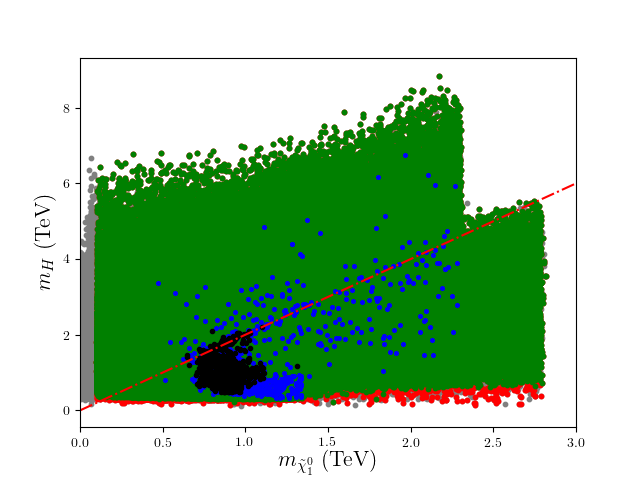} 
 	\caption{The mass spectrum of the lightest neutralino and chargino and relic density channels over the following planes:  $(m_{\tilde{\chi}_{1}^{0}},m_{\tilde{\chi}_{1}^{\pm}})$ (top left), 
 		$(m_{\tilde{\chi}_{1}^{0}},m_{\tilde{\tau}})$ (top right), 
 		$(m_{\tilde{\chi}_{1}^{0}},m_A)$ (bottom left) and $(m_{\tilde{\chi}_{1}^{0}},m_H)$ (bottom right). Our colour convention is as listed at the end of Section \ref{sec:scan}.}
 	\label{fig:neutralino}
 \end{figure}

 In Fig. \ref{fig:neutralino} we illustrate the neutralino and chargino mass spectrum and relic density channels in $(m_{\tilde{\chi}_{1}^{0}},m_{\tilde{\chi}_{1}^{\pm}})$ (top left), $(m_{\tilde{\chi}_{1}^{0}},m_{\tilde{\tau}})$ (top right),  $(m_{\tilde{\chi}_{1}^{0}},m_A)$ (bottom left) and $(m_{\tilde{\chi}_{1}^{0}},m_H)$ (bottom right) planes. The colour coding is the same as in Fig. \ref{fig:spar}. In top panels, the diagonal solid red lines indicate regions in which the displayed parameters are degenerate in value. The top left panel shows that the lightest neutralino, which are our DM candidate, versus the lightest chargino. The mass of DM consistent with Yukawa unification can be in a narrow range as $ 0.6$ TeV $\lesssim m_{\tilde{\chi}_{1}^{0}} \lesssim 1.3 $ TeV while the lightest chargino masses are also squezed to the range of  $ 0.8$ TeV $\lesssim m_{\tilde{\chi}_{1}^{\pm}} \lesssim 1.5 $ TeV. Furthermore, as can be seen from the top left panel, the lightest chargino and LSP can not be degenerate in mass. This means that the solutions which corresponds to chargino-neutralino coannihilation channels are not sufficient to reduce the relic abundance of the LSP to correct relic density. It is concluded that our LSP can not be higgsino-like since our heavier charginos are mainly composed of higgsinos. Top right panel shows that the stau-neutralino coannihilation channels are almost ruled out in the case of the YU except for a few number of solution. The bottom panels depict LSP versus the masses of CP-odd  and second lightest CP-even Higgs. The solid red line shows the points with $m_{A,H}=2m_{\tilde{\chi}_1^{0}}$, condition on setting the dominant resonant DM annihilation via A or H mediation. According to the plots, the $A(H)$ resonance solutions with $ 0.5$ TeV $\lesssim m_{\tilde{\chi}_{1}^{0}} \lesssim 2 $ TeV are possible in the all mass range of the LSP which are consistent with the YU.
 
\begin{figure}[!t]
	\centering
	\includegraphics[width=0.48\linewidth]{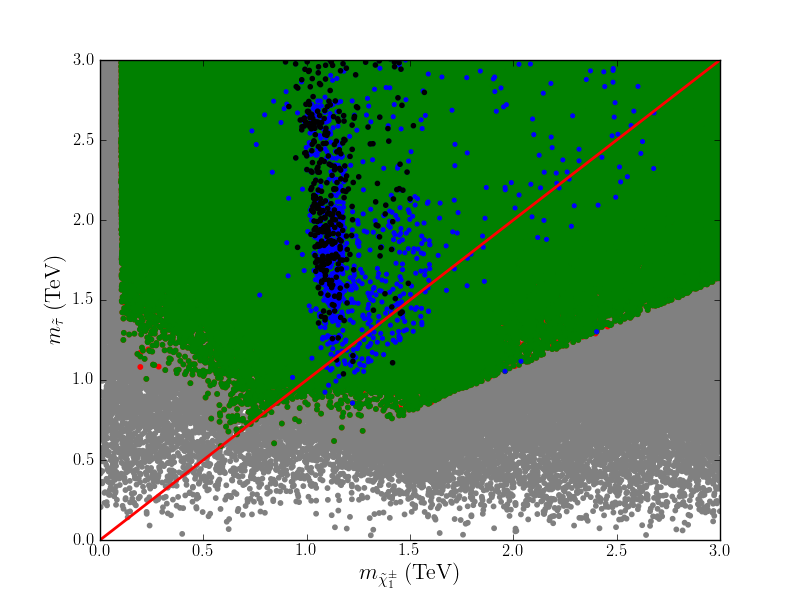}
	\includegraphics[width=0.48\linewidth]{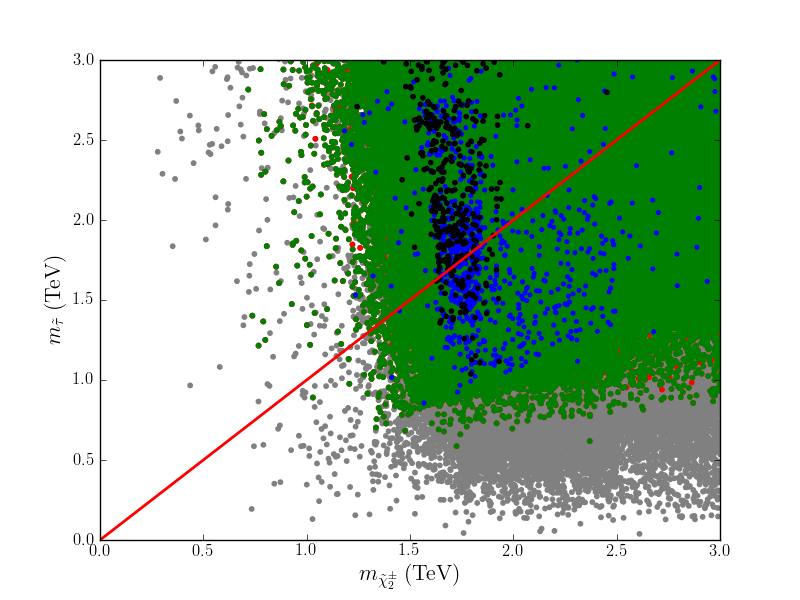} 
	\caption{The mass spectrum of the chargino versus stau over the following planes:  $(m_{\tilde{\chi}_{1}^{\pm}},m_{\tilde{\tau}})$ (left), 
		$(m_{\tilde{\chi}_{2}^{\pm}},m_{\tilde{\tau}})$ (right). Our colour convention is as listed at the end of Section \ref{sec:scan}.}
	\label{fig:chargino}
\end{figure}

The recent LHC analyses have excluded the chargino masses for $ m_{\tilde{\chi}_{i}^{\pm}} < 1.1 $ TeV in the chargino-neutralino pair production channels with decays via sleptons \cite{Sirunyan:2017lae}, where i$  = 1,2 $ stand for the lighter and heavier chargino respectively. In Fig. \ref{fig:chargino}, we present the mass relations between the charginos and stau. The diagonal solid red lines indicate regions in which the displayed masses are degenerate in value. As seen from the figure, the solutions favourable to YU below red lines, where the charginos can decay to stau kinematically, can survive from the aforementioned LHC bounds for the chargino masses and they are reachable at the LHC Run-III. 

\begin{figure}[!t]
	\centering
	\includegraphics[width=0.48\linewidth]{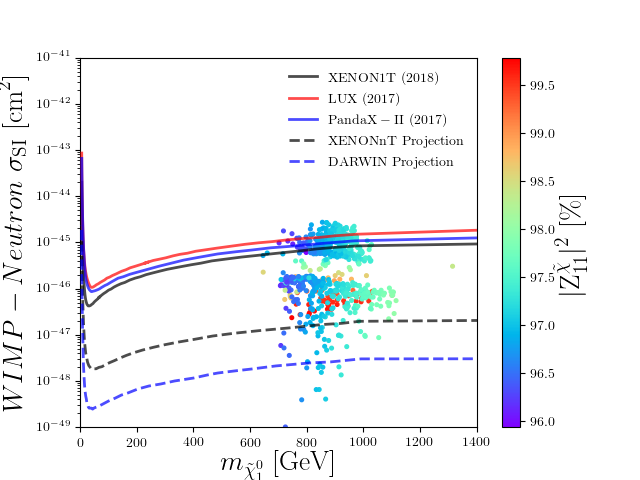}
	\includegraphics[width=0.48\linewidth]{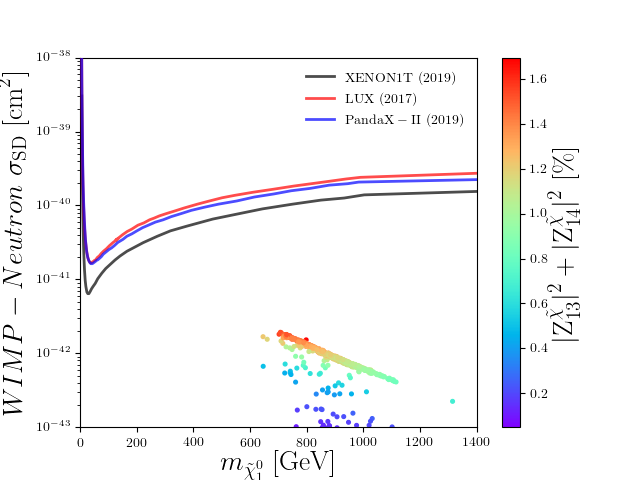}
	\caption{DM-neutron SI (left) and  SD (right) scattering cross section  as a function of the  mass of the lightest neutralino LSP. The colour bars show the composition of the LSP. Limits from current (solid) and future (dashed) experiments are also shown.}
	\label{fig:SI_SD}
\end{figure}

Before concluding, we also present the status of the NHSSM solutions that are compatible with $ t-b-\tau $ YU in the recent and future DM direct detection searches. In Fig. \ref{fig:SI_SD} shows the DM-neutron Spin-Independent (SI, left panel) and  Spin-Dependent (SD, right panel) scattering cross sections  as  functions of the the neutralino LSP . The colour bars show the bino (the left panel) and higgsino(the right panel) compositions of the DM.  Here, the points satisfy the $ t-b-\tau $ YU and all the experimental constraints used in this work. They correspond to the ''Black'' points as described at the end of  Section \ref{sec:scan}. As seen from the colour bars in the both side of the figure, all $ t-b-\tau $ Yukawa unified solutions include bino-like LSP neutralinos with a very small higgsino percentage. In the left panel, the black, blue and red solid lines show XENON1T \cite{Aprile:2018dbl}, PandaX-II \cite{Cui:2017nnn} and LUX \cite{Akerib:2016vxi} upper limits for the  SI  ${\tilde\chi}_1^0 $ - n cross section, respectively, while the black and blue dashed lines illustrate the prospects of the XENONnT and DARWIN for future experiments \cite{Aalbers:2016jon}, respectively. As seen from this panel, most of our points with the LSP mass in the range of $ 0.6$ TeV $\lesssim m_{\tilde{\chi}_{1}^{0}} \lesssim 1.3 $ TeV are presently consistent with all direct detection experimental constraints and can be probed by the  next generation of experiments for SI cross section. In the right panel, the black, blue and red solid lines show XENON1T \cite{Aprile:2019dbj}, PandaX-II \cite{Xia:2018qgs} and LUX \cite{Akerib:2017kat} upper limits for the SD ${\tilde\chi}_1^0 $ - n cross section, respectively. As seen from this plot, all solutions are consistent with current experimental SD direct detection results for all the LSP masses.
\section{Conclusion}
\label{sec:conc}

In this paper, we have studied the $ t-b-\tau $ Yukawa unification in the non-holomorphic MSSM with universal boundary conditions  and explored the low scale and DM implications in this scenario. We have scanned the parameter space such that the LSP is always the lightest neutralino ${\tilde\chi}_1^0$. Following this, we have applied all current collider and DM bounds onto the parameter space of the model. Additionally, we demand Yukawa unification at the grand unification scale ($M_{{\rm GUT}}$). The YU condition, $ R_{tb\tau} \leq 1.1 $ is achieved and the fundamental parameters of NHSSM are found to be in a range such as $m_{0} \gtrsim 2$ TeV, $ 1.5 \ {\rm TeV} \lesssim M_{1/2} \lesssim 3 \ {\rm TeV}$. The $\tan\beta $ parameter is mostly restricted to the region in the interval $ [46,55] $ by the YU condition. Also, YU strictly requires the negative and large values for the NH terms $ A_0^\prime $ and $ \mu^prime $ as $ -15 \ {\rm TeV} \lesssim A_0^{\prime} \lesssim -8 \ {\rm TeV}$ and $ \mu^\prime < -4.5 \ {\rm TeV} $. We find that the mass of the coloured scalar particles  are in the range of $ 3 \ {\rm TeV} \lesssim m_{\tilde{b}},m_{\tilde{t}} \lesssim 5 \ {\rm TeV}$ while the stau $ \tilde{\tau} $ can be as light as 1 TeV. We also obtain the gluino mass with $ m_{\tilde{g}} \gtrsim 3 \ {\rm TeV} $. Furthermore, we find the chargino solutions in which the charginos can decay to stau kinematically and the YU can partly be tested in this channel at the LHC Run III.

 In the DM sector, the solutions consistent with all current experimental bounds coming from relic density and direct detection experiments were found  a bino-like LSP neutralino with $ 0.6$ TeV $ \lesssim m_{\chi_{1}^{0}} \lesssim 1.3 $ TeV. In this respect, we have been able to identify only $A$ (the pseudoscalar Higgs state) mediated resonant annihilation as the main channel rendering our DM scenario consistent with Planck measurements. Furthermore, as for SI and SD ${\tilde\chi_1^0}$ - n scattering cross section bounds from DM direct detection experiments, we have seen that YU scenarios are mostly viable (compliant with present limits) yet they could be detected by the next generation of such experiments.

\acknowledgments

The author thanks Levent Solmaz and Cem Salih Un for valuable discussions. The work of YH is supported by The Scientific and Technological Research Council of Turkey (TUBITAK) in the framework of  2219-International Postdoctoral Research Fellowship Program. The author also acknowledge the use of the IRIDIS High Performance Computing Facility, and associated support services at the University of Southampton, in the completion of this work.






\bibliography{YU_NHSSM_JHEP.bib}

\end{document}